\documentclass[conference]{IEEEtran}

\IEEEoverridecommandlockouts
\usepackage{cite}
\usepackage{amsmath,amssymb,amsfonts}
\usepackage{algorithmic}
\usepackage{graphicx}
\usepackage{textcomp}
\usepackage{xcolor}
\def\BibTeX{{\rm B\kern-.05em{\sc i\kern-.025em b}\kern-.08em
    T\kern-.1667em\lower.7ex\hbox{E}\kern-.125emX}}
\begin{document}

\title{Fine-grained Early Frequency Attention for Deep Speaker Recognition\\
\thanks{The authors would like to thank IMRSV Data Labs for their support of this work. The authors would also like to acknowledge the Natural Sciences and Engineering Research Council of Canada (NSERC) for supporting this research (grant number: CRDPJ 533919-18).}
}

\author{\IEEEauthorblockN{Amirhossein Hajavi, Ali Etemad}
\IEEEauthorblockA{\textit{Dept. ECE and Ingenuity Labs Research Institute} \\
\textit{Queen's University}\\
Kingston, Canada \\
\{a.hajavi, ali.etemad\}@queensu.ca}
}

\maketitle
\begin{abstract}

Attention mechanisms have emerged as important tools that boost the performance of deep models by allowing them to focus on key parts of learned embeddings. However, current attention mechanisms used in speaker recognition tasks fail to consider fine-grained information items such as frequency bins in input spectral representations used by the deep networks. To address this issue, we propose the novel Fine-grained Early Frequency Attention (FEFA) for speaker recognition in-the-wild. Once integrated into a deep neural network, our proposed mechanism works by obtaining queries from early layers of the network and generating learnable weights to attend to information items as small as the frequency bins in the input spectral representations. To evaluate the performance of FEFA, we use several well-known deep models as backbone networks and integrate our attention module in their pipelines. The overall performance of these networks (with and without FEFA) are evaluated on the VoxCeleb1 dataset, where we observe considerable improvements when FEFA is used.
\end{abstract}
\section{Introduction}
Extracting deep representations from audio signals has been the subject of a large number of studies. Many deep learning techniques have recently been used to extract deep representations in speech-related tasks such as speech recognition \cite{schneider2019wav2vec}, speech disfulency detection \cite{kourkounakis2020fluentnet, kourkounakis2020detecting}, and speaker recognition (SR) \cite{hajavi2019deep, chung_voxceleb, xie2019utterance, hajavi2021siamese, ghahabi_deep_2017, kim_deep_2019, yu_ensemble_2019}. Many techniques have been developed using Convolutional Neural Networks (CNNs) \cite{hajavi2019deep, okabe_attentive_2018, xie2019utterance, chung_voxceleb, nagrani_VoxCeleb} for performing SR on utterances with different domains such as telephone conversations or in-the-wild scenarios.  

To learn audio signals with deep CNNs, they are typically first transformed to the time-frequency domain using techniques such as Short-term Fourier Transform (STFT). The resulting transformations, namely spectrograms, are then passed to a CNN as input for extraction of deep speech representations. In audio representation learning, specific frequency bins in the spectrogram representations often contain significant task-specific information \cite{hansen_speaker_2015}. As a result, to effectively learn the prominent task-specific features through deep neural pipelines, particular frequency bins in the input representations need to be accentuated as they are entered into the model.



Attention models have been used in many domains \cite{hardik2020attention, patrick2019capsule, patrick_LSTM_ATTENTION} to determine the importance of information items in extracted representations of input data. However, current attentive models in SR only operate on latent embeddings extracted from deep layers of deep neural networks (DNNs). Meanwhile, DNNs used to learn audio representations typically perform non-linear operations on the input frequency bins. This makes the task of determining the importance of each individual frequency bin quite complex. Therefore we propose two properties for an effective attention mechanism in SR: (1) focusing on individual frequency bins in the input spectral representation (i.e. the mechanism should be \textit{fine-grained}); (2) focusing on input spectrogams that have not been heavily modified by deep layers of the model (i.e. the mechanism should be applied \textit{early} in the pipeline). 

In this paper, to address the requirements mentioned above for deep speaker recognition in-the-wild, we introduce the \textit{fine-grained early frequency attention (FEFA)}. Our method is integrated prior to the early layers of DNN pipelines, and operates on the input spectral representations early on in the DNNs. This enables the network to attend to information items as small as frequency bins. We evaluate our model using three well-known CNN architectures against the VoxCeleb evaluation set. The results show that by coupling our attention model with these architectures, the performance of these networks are considerably improved. 

Our contributions in this paper can be summarized as follows: (\textbf{1}) We introduce a novel attention mechanism to help extract better deep embeddings of utterances that contain speaker-related information; (\textbf{2}) We evaluate our method on the task of SR using the widely-used VoxCeleb dataset and demonstrate considerable improvements when used with different DNNs; (\textbf{3}) We also test the robustness of our model against added synthetic noise and observe improvements in robustness compared to the non-attentive baselines.


\section{Related work}


Attentive models have been widely used in SR. Earlier models proposed in \cite{bhattacharya2017deep} and \cite{rahman2018attention} utilized fully connected (FC) layers for the attention mechanism. In \cite{bhattacharya2017deep}, features extracted from utterances using a CNN with an architecture similar to VGGnet \cite{simonyan2014very}, were given to the attention model to create an enhanced embedding of the utterance. In \cite{rahman2018attention}, an RNN was used to extract the first set of features from utterances. The same attention model was then used to extract the utterance-level embeddings. In \cite{safari2019self}, a multi-head attention mechanism was used to attend to features extracted by a CNN.

The methods proposed in \cite{okabe_attentive_2018, zhu_self-attentive_2018} used Time-Delay Neural Networks (TDNN) for extracting features from input utterances. Attention mechanisms were then used to aggregate frame-level features to form utterance-level embeddings. These embeddings were then used for SR. These approaches were evaluated against the NistSRE16 in \cite{zhu_self-attentive_2018} and VoxCeleb evaluation set in \cite{okabe_attentive_2018}, showing improvements in performance compared to baseline models. 

\begin{figure*}[t]
\centering
\includegraphics[width=0.7\linewidth]{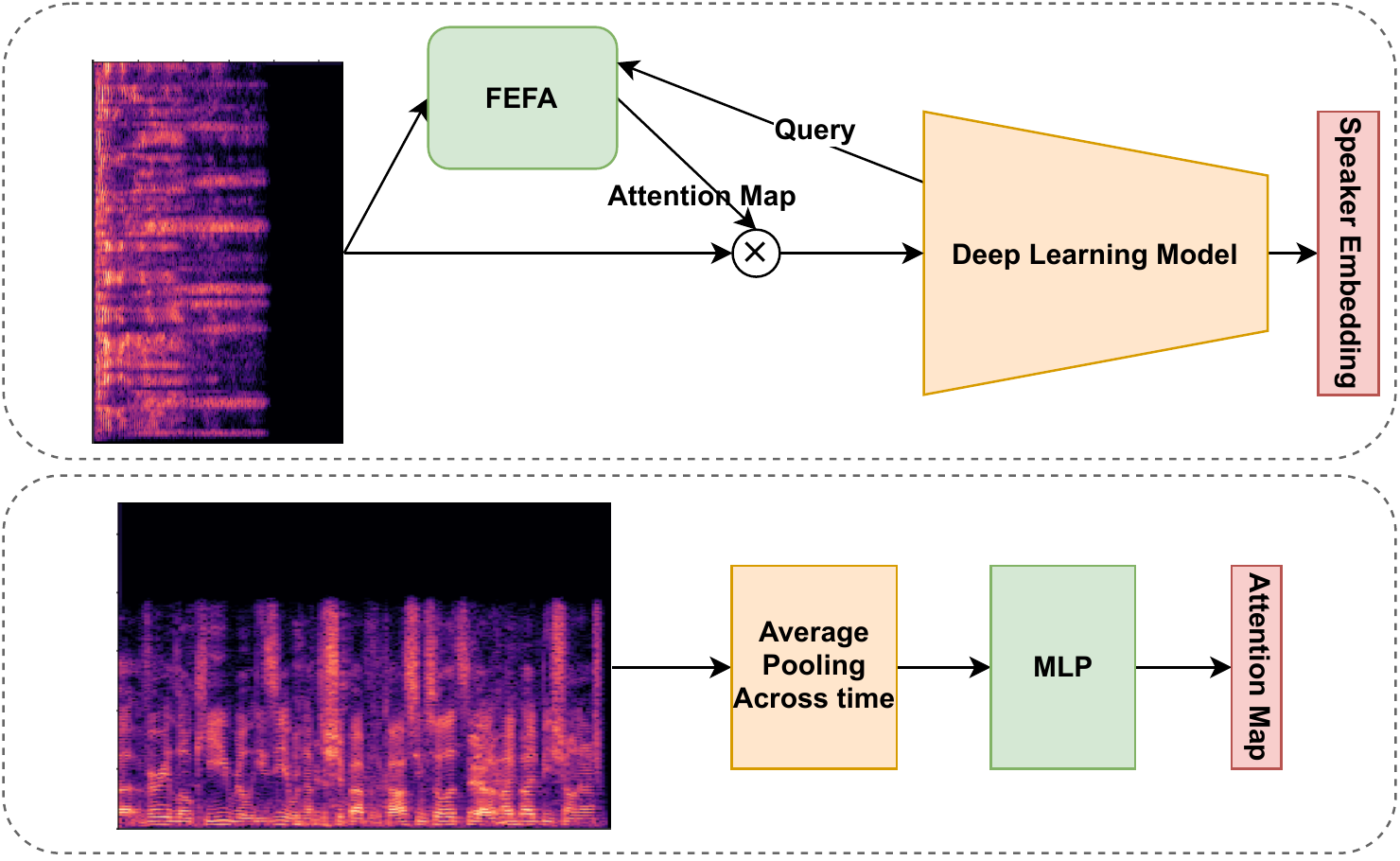}
\caption{\small An overview of the FEFA (top) and its internal mechanism (bottom) are illustrated.}
\label{fig:ffa_scheme}
\end{figure*}

In \cite{bian2019self}, a different attention mechanism was explored for SR, where a CNN-based self-attentive network was proposed that used a ResNet backbone to extract features from reference and test utterances. 
The solution was shown to achieve a higher accuracy than baseline ResNet models. In \cite{zhang_seq2seq_2019}, an attention model was used to focus on discriminative features of the obtained embeddings of two utterances. The results acquired from their architecture showed significant improvements compared to baseline models when performing text-dependent speaker verification. In \cite{kye2021supervised}, a supervised attention model was used to obtain utterance embeddings with more discriminative information compared to the benchmarks. In a different approach taken in \cite{yadav2020frequency}, an early attention mechanism was proposed using convolutional block attention modules \cite{woo2018cbam}. In their proposed method, attention blocks consisting of shared-weight CNN kernels were used to attend to the input spectrogram, showing a considerable improvement in the performance compared to the previous models. 

In the majority of aforementioned studies, the proposed attention mechanisms are applied on latent embeddings of the utterances or use mechanisms with shared weights to attend to the input spectrograms. This prevents the attention model from focusing on granular features such as frequency bins. In this work we aim to employ a fine-grained attention mechanism early in the DNN pipeline so that granular task-specific items in the input spectrograms can be accentuated.


\section{Method}

\subsection{Proposed Approach}
DNN models used to learn speech representations generally exploit spectrograms as inputs, which are extracted by a non-learnable process. Therefore the resulting spectrograms do not possess the means to differentiate between important frequency bins and unimportant ones with respect to the task at hand. Let's assume $E$ is a deep embedding extracted from input spectrogram $X$, by a DNN $\mathcal{F}$ as $E = \mathcal{F}(X)$. The non-linearity of $E$ prevents these attention mechanisms from determining a unique level of contribution for each frequency bin throughout the embedding. To address this issue, the attention needs to be employed prior to input $X$ being processed by the deep model. In other words, $X$ should first be enhanced through an attention model $\mathcal{A}$ by $E = \mathcal{F}(\mathcal{A}(X))$.

We propose fine-grained early frequency attention (FEFA) that operates on spectral representations of utterances prior to DNN representation learning. Our FEFA model is designed to attend to each frequency bin individually and asses its level of contribution via the returned gradients from early layers of the DNN. Our model works by enhancing the input through applying a learned attention map \textit{$\mathcal{M}$} over $X$ the input spectrogram of the utterance. The overall scheme of the model is shown in Figure \ref{fig:ffa_scheme} (top). We take advantage of early layers of the DNN model as the source of the attention queries to help the model to localize the attention map for each frequency bin, resulting in enhanced performance. Additionally, to enhance each single bin, our approach relies on fewer parameters compared to non-localized approaches, resulting in low complexity of the mechanism.

Figure \ref{fig:ffa_scheme} (bottom) shows the internal scheme of our FEFA model. An average pooling operation is first performed on the spectrogram representation of the utterance over the \textit{time} axis. The resulting vector is then passed through the kernel of the attention model. The kernel determines the impact of each frequency bin in the form of probabilities within the improved spectrogram representation of the utterance using:
\begin{equation}
  p_i = e^{I(x_i)\times W}/\sum\limits_{j=1}^{|M|} e^{I(x_j)\times W} ,
  \label{eq:FFA_1}
\end{equation}
where $I(x_i)$ is the one-hot encoding representation of the index of the frequency bin in the input vector, and $W$ is the learnable weight vector of the kernel. The probabilities calculated through this process are then used to create the corresponding index $i$ in the attention map $\mathcal{M}$ for frequency bin $x_i$. The attention map is then calculated using $\mathcal{M}(i) = p_i \cdot x_i $.
The resulting attention map is then applied to the spectral representation of the utterance leading to an improved spectrogram to be passed to the DNN model for representation learning and subsequent classification/regression. 

The FEFA model, in contrast to the typical attention models, does not require a deep embedding of utterances to operate on, and is therefore not dependant on the architecture of the deep model used for SR. As a result, FEFA can be attached to various deep learning architectures that utilize spectral representations of utterances as input. In this paper we implement the proposed FEFA model on various backbone DNNs namely ResNet, VGG, and SEResNet to show the performance gain using this model with various architectures.

The kernel of the attention module consists of a single layer of locally connected MLP. We select a simple kernel in order to ensure that any performance gain has been obtained through the use of our early attention mechanism only and not by the complex kernels used in the process. Additionally, the sparse connections integrated in this type of MLP enables us to localize the attention map towards each frequency bin. 



\subsection{Backbone Networks}
We have selected three DNN models as backbone networks to assess the impact of the FEFA model. We use two of the recently introduced DNN models based on VGGNet \cite{nagrani_VoxCeleb} and ResNet \cite{xie2019utterance} as our benchmarks. The details of the architecture of these models are borrowed from their respective papers. We also introduce a new thin-SEResNet benchmark model by combining the ResNet-based model mentioned above with the SE blocks proposed in \cite{hu2018squeeze}. We evaluate these benchmark models in three different settings: 1) The bare backbone model without the use of FEFA. 2) The backbone model accompanied with a single layer of FEFA placed before the input layer. 3) The model with FEFA integrated in multiple depths of the DNN pipeline wherever the hidden time-frequency representation changes in size.



For the last backbone network we have implemented a new thin-SEResNet model with inspirations from the thin-ResNet model proposed in \cite{xie2019utterance} and the technique proposed in \cite{hu2018squeeze}. The overall architecture of the thin-SEResNet model is similar to the thin-ResNet model with the addition of Squeeze-and-Excitation (SE) modules inside the residual blocks. The SE module first employs a global pooling layer which accumulates the information across each channel. The output of the pooling layer is then passed through two FC layers followed by two activation functions of ReLU and Sigmoid. The result, which is a vector with a size equal to the number of channels, is then multiplied over the channels of the embedding in the main pipeline of the residual block.



\subsection{Training} 
In order to train our model, we extract spectrogram representations of utterances with 257 frequency bins. The number of frequency bins used in this experiment is selected to match the input of the baseline models provided by the authors in \cite{nagrani_VoxCeleb, xie2019utterance}. Using the same input as that of the previous works, along with similar architectures for backbone networks, enables us to perform accurate evaluations knowing that the performance gain is solely obtained via the use of our FEFA module. Accordingly, we use the frame length of 25 \textit{ms} with overlap of 10 \textit{ms} while extracting spectrogram representations. We then train the three settings (without the FEFA module, a single integration, and multiple integrations) of our model on each backbone network using the softmax loss function. We use the initial learning rate of 0.0001. We have selected this learning rate approach to reduce the chance of the models being trapped in local minima. Training was done with a batch size of 64, on a single Nvidia Titan RTX (24 GB VRAM) GPU.

\section{Experiments and Results}

\subsection{Dataset}
All the implementations of FEFA are trained using in-the-wild dataset VoxCeleb2 \cite{chung_voxceleb} and tested against the standard version of VoxCeleb1 \cite{nagrani_VoxCeleb} evaluation set in order to better compare the performance of our model against prior works such as \cite{okabe_attentive_2018,nagrani_VoxCeleb, zhu_self-attentive_2018}.

\begin{table*}[!t]
\footnotesize
\centering
\caption{The results for performing speaker recognition using backbone models with and without integration of FEFA. *The original work in \cite{yadav2020frequency} reports the EER values with ArcSoftmax loss. However, in order to maintain consistency and provide a fair comparison, we retrained the model with Softmax.}
\resizebox{1\textwidth}{!}
{%
\begin{tabular}{l|c|c|c|c|c|c|c}
\hline
study                                             & Backbone      &  Loss      & Attention              & Dims      & Aggregation      & Training Set      & EER (\%)\\ \hline \hline
Nagrani et al. \cite{nagrani_VoxCeleb}             & i-Vector+PLDA & Softmax & --                     & --        & --               & VoxCeleb1         & 8.8     \\
Nagrani et al. \cite{nagrani_VoxCeleb}             & VGG           & Softmax  & --                     & 1024      & TAP              & VoxCeleb1         & 7.8     \\
Hajibabai et. \cite{hajibabaei2018unified}        & ResNet20      & Softmax  & --                     & 128       & TAP              & VoxCeleb1         & 4.30    \\
Chung et al.  \cite{chung_voxceleb}               & ResNet50      & Softmax  & --                     & 512       & TAP              & VoxCeleb2         & 3.95    \\
Xie et al. \cite{xie2019utterance}                & Thin-ResNet   & Softmax  & --                     & 512       & TAP              & VoxCeleb2         & 10.48   \\
Xie et al. \cite{xie2019utterance}                & Thin-ResNet   & Softmax  & --                     & 512       & GhostVLAD        & VoxCeleb2         & 3.22    \\
Okabe et al. \cite{okabe_attentive_2018}          & x-Vector      & Softmax  & Soft Attention         & 1500      & AP              & VoxCeleb1         & 3.85    \\
Bian et al. \cite{bian2019self}                   & ResNet50      & Softmax  & Self Attention         & 512       & AP                          & VoxCeleb2          & 5.4     \\
Kye et al. \cite{kye2021supervised}               & Resnet34      & Softmax  & Supervised Attention   & 256       & AP        & VoxCeleb2          & 4.75    \\
Yadav et al.* \cite{yadav2020frequency}      & Thin-ResNet         & Softmax*  & CBAM                   & 512       & GhostVLAD        & VoxCeleb2          & 3.10    \\ 
\hline

Proposed                                          & VGG       & Softmax       & FEFA(M)               & 1024      & TAP              & VoxCeleb2           & 7.80   \\
Proposed                                          & VGG        & Softmax      & FEFA(S)               & 1024      & TAP              & VoxCeleb2           & 7.40   \\
Proposed                                          & SEResNet   & Softmax      & FEFA(M)               & 512       & TAP              & VoxCeleb2           & 4.81   \\
Proposed                                          & SEResNet   & Softmax      & FEFA(S)               & 512       & TAP              & VoxCeleb2           & 4.58   \\ 
Proposed                                          & Thin-ResNet   & Softmax   & FEFA(M)               & 512       & TAP              & VoxCeleb2           & 5.32   \\
Proposed                                          & Thin-ResNet   & Softmax   & FEFA(S)               & 512       & TAP              & VoxCeleb2           & 5.40   \\
Proposed                                          & Thin-ResNet   & Softmax   & FEFA(M)               & 512       & GhostVLAD        & VoxCeleb2           & 3.18   \\
Proposed                                          & Thin-ResNet   & Softmax   & FEFA(S)               & 512       & GhostVLAD        & VoxCeleb2           & \textbf{2.85}   \\

\hline
\end{tabular}
}
\label{tab:results}
\end{table*}

\begin{figure*}[t]
\centering
\includegraphics[width=.85\linewidth]{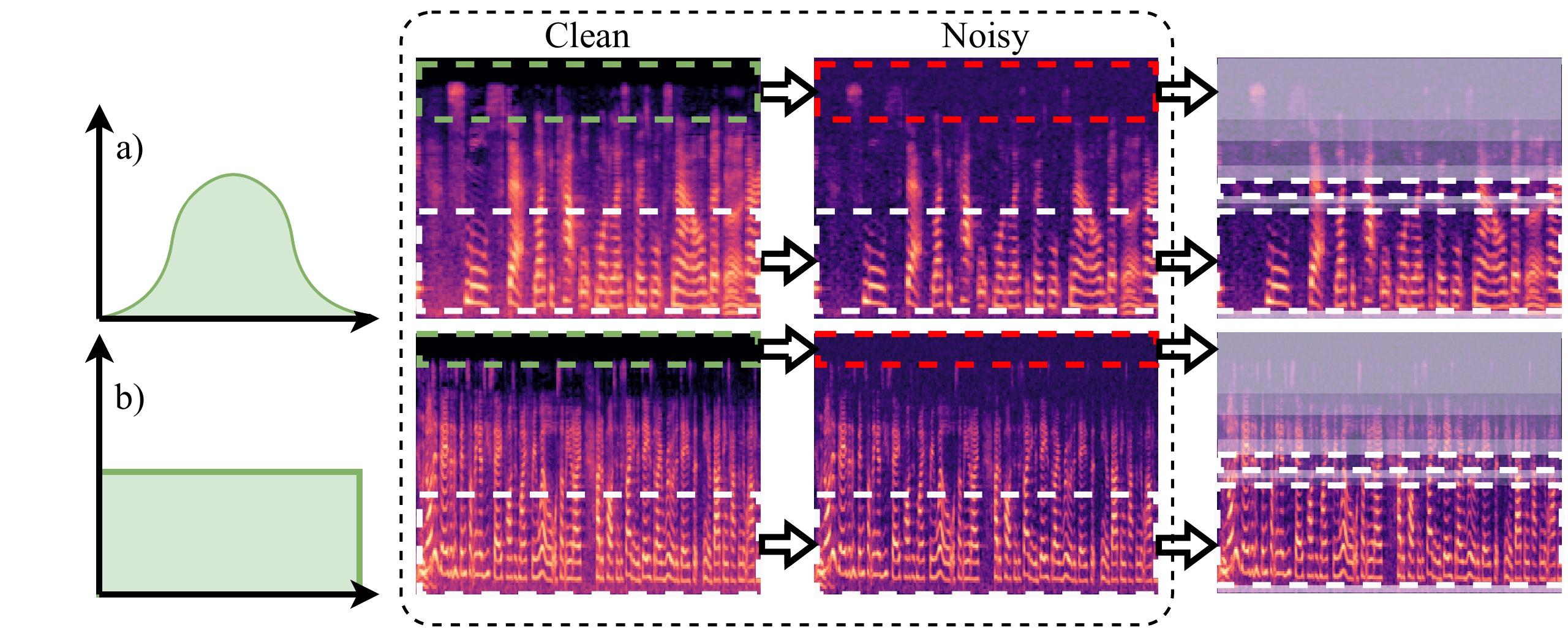}\caption{\small Robustness experiments by adding synthetic noise to input utterances, where noise is selected from a Gaussian distribution (a) and uniform distribution (b). The green and red rectangles denote areas most affected by noise. In the rightmost column, the attention map generated by the FEFA model is overlaid on the spectrogram. }
\label{fig:spectrogram_samples}
\end{figure*}

\subsection{Performance}

We evaluate our proposed attention mechanism using Equal Error Rate (EER) to measure the performance of the models in speaker verification. The results are presented in Table \ref{tab:results} along with the performance gain ($\Delta$\textit{EER}) achieved by using FEFA. It is shown that by adding the FEFA to the backbone networks, a performance gain of 3.1\% has been achieved. 

The results of our experiments with VGG and thin-ResNet backbones are compared to the results reported directly in \cite{nagrani_VoxCeleb, xie2019utterance}. However no such study could be found for the use of thin-SEResNet. Therefore the implementation and evaluation of the thin-SEResNet model as a backbone network is done in the course of this experiment. As the results suggest, the FEFA model has a positive impact on the performance of all the backbone models regardless of the architecture. 

We also compare our work to previous studies that use conventional attention mechanisms for SR, and present the results in Table \ref{tab:results}. Here, \cite{bian2019self} and \cite{kye2021supervised} have used self-attention and supervised attention, respectively. In the original baseline papers, various loss functions have been tested. In this table, however, we only report the results obtained by `softmax' loss to remain consistent with the other benchmarks.  
Moreover, we also compare our model with the early frequency attention approach proposed in \cite{yadav2020frequency}, where we train the model using softmax loss. The results in the table demonstrate that FEFA, when integrated into the different backbones (VGG, Thin-ResNet, SE-ResNet), considerably improves SR results and outperforms the original non-attentive baselines.


As an interesting observation, it can be seen that although the use of FEFA integrated in multiple depths of the DNN pipeline improves the performance of the backbone models by a considerable margin, the performance gain is not as much as a single integration of the FEFA module. This may be due to the fact that the use of 2D convolutional kernels in the DNN models results in features from the time axis to be convolved with the features residing in the frequency axis; hence reducing the contribution of the weighted frequency bins in the resulting representation.  




\begin{figure*}[t]
\centering
\includegraphics[width=.85\linewidth]{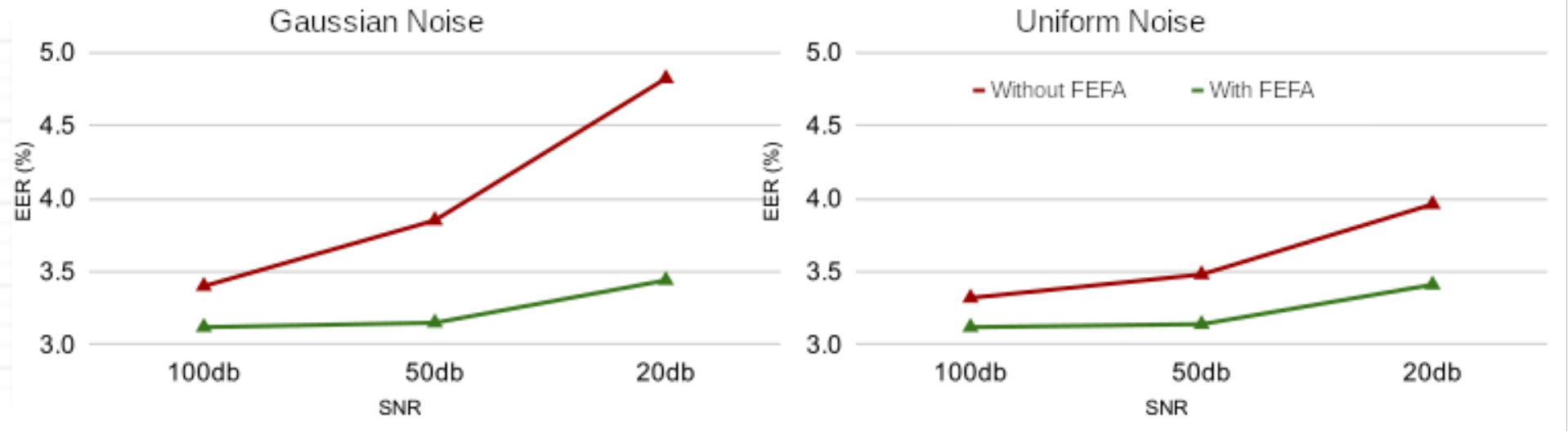}
\caption{\small Results of the robustness experiments where synthetic noise is added to input utterances. The comparison is performed with a Thin-ResNet model, with and without FEFA.}
\label{fig:ROBUSTRESULTS}
\end{figure*}

\subsection{Robustness to Noise}
We extend our experiments by assessing the performance of the backbone models equipped with FEFA against different levels of added noise. Given the in-the-wild nature of the VoxCeleb dataset and presence of natural noise in the dataset, adding noise from other \textit{real} sources may not necessarily introduce a new set of challenges for the model. Therefore, we test the performance of the proposed model against the standard evaluation set of VoxCeleb by adding different levels of \textit{synthetic} noise. In order to test the effectiveness of our FEFA model against noise, we select the best performing setup from previous experiments, namely the Thin-ResNet, and compare its performance with the same backbone model when FEFA is added.

We select two distributions, namely Gaussian (Figure \ref{fig:spectrogram_samples} (a)) and uniform (Figure \ref{fig:spectrogram_samples} (b)), from which we sample for generating the synthetic noise. The synthetic noise is added to the test utterances prior to spectrogram calculation. The effect of the added noise can be seen by comparing the clean and noisy spectrograms in Figure \ref{fig:spectrogram_samples}. The areas most affected by artifact noise are highlighted by green rectangles in the clean spectrograms and the red rectangles in the noisy spectrograms. In the rightmost column of the figure, the attention map generated by the FEFA model is overlaid on the spectrogram. As shown in the figure the areas most affected by the artifact noise have the least weights in the attention map, hence their contribution to the final learned representation is decreased. On the other hand, the parts of the spectrogram with the most amount of information (shown in the figure with the white dotted rectangles) are given higher weights, which boosts their contribution towards calculation of the final embedding.




Figure \ref{fig:ROBUSTRESULTS} reports the results of the above-mentioned experiment. We evaluate the performance of the model against the VoxCeleb evaluation set and add the synthetic noise with 3 signal-to-noise ratios (SNR) of 20db, 50db, and 100db. The difference in the performance of the models are illustrated in the figure where our model (depicted by green lines) shows more robustness compared to the baseline \cite{xie2019utterance} (depicted by red lines). We observe that the backbone networks suffer from a large drop in performance when the noise is added, while the model equipped with FEFA is not affected as much. This shows that the models equipped with the FEFA module are relatively more robust compared to their counterparts.

\section{Summary and Future Work}
This study introduces a novel attention mechanism, FEFA, for speaker recognition. Our proposed attention enhances the spectral representations of utterances by attending to individual frequency bins. The results of our experiments show considerable improvements in the performance of various DNN models. Our experiments on the robustness of the backbone networks with FEFA against added synthetic noise further highlights the benefits of our proposed attention module. For future steps of this study we intend to use more sophisticated internal components, as well as more advanced temporal pooling, to help achieve higher performance gains.

\bibliographystyle{IEEE}
\bibliography{references}
\end{document}